\documentclass{article}

\usepackage{graphicx} % Required for inserting images
\usepackage[a4paper,total={17.5cm,24cm},top=1.5cm, bottom=1.8cm, left=1.5cm, right=1.5cm]{geometry}

\usepackage{color}
\usepackage{xcolor}
\usepackage{soul}

\usepackage{authblk}

\usepackage[
backend=biber,
style=numeric,
sorting=none
]{biblatex} %Imports biblatex package
\title{Thermal tuning of dynamic response in Ag-based nanowire networks}

\author[1]{J.I. Díaz Schneider}
\author[2]{C. Gomez}
\author[3,4]{C. Acha}
\author[1]{P.E. Levy}
\author[1]{E.D. Martínez}
\author[5]{C.P. Quinteros}

\affil[1]{Consejo Nacional de Investigaciones Científicas y Técnicas (CONICET), Argentina.\\
Instituto de Nanociencia y Nanotecnología (CNEA - CONICET), Nodo Bariloche.\\
Gerencia Física, Centro Atómico Bariloche, Comisión Nacional de Energía Atómica (CNEA), Av. Bustillo 9500, (8400) S. C. de Bariloche, Río Negro, Argentina.
}%
\affil[2]{Instituto Prof. Jorge Sábato (UNSAM-CNEA), San Martín (1650), Argentina.}
\affil[3]{Universidad de Buenos Aires, Facultad de Ciencias Exactas y Naturales, Departamento de Física. Buenos Aires, Argentina.}
\affil[4]{CONICET - Universidad de Buenos Aires, Instituto de Física de Buenos Aires (IFIBA). Buenos Aires, Argentina.}
\affil[4]{%
Instituto de Ciencias Físicas (ICIFI, UNSAM-CONICET), Martín de Irigoyen 3100, San Martín (1650), Argentina.
}%

\date{\today}

\begin{document}

\maketitle

\begin{abstract}
    \noindent 
    %Self-assembled networks of metallic nanowires (NW) are being intensively explored as test benches for neuromorphic proposals. In this work, we use external temperature to: i. characterize dense assemblies of AgNW and ii. affect the population of metallic junctions, unveiling a rich diversity of dynamic responses. By using impedance spectroscopy, the metallic nature of the pristine low-resistive state is credited. Thermal annealing triggers the achievement of a capacitive response, which is successfully rationalized using a previously developed model. Modulating the external temperature reveals an indirect strategy to effectively modify the water content at the NWs’ intersections and, consequently, enable multiple switching schemes suitable for brain-like processing alternatives. 
    Self-assembled networks of metallic nanowires (NWs) are being intensively explored as test benches for neuromorphic proposals.  In this work, we study the electric transport properties of dense self-assembled networks of Ag-based NWs (AgNWNs) coated with a thin insulating layer, using DC and AC stimuli. The building blocks of this network are the metallic NWs and the NW-NW junctions, either metallic or memristive. In the pristine state, frequency independence of the impedance reveals an over-percolated purely resistive network. A combination of low-temperature annealing and AC stimulus is shown to drastically affect the resistivity of the sample (interpreted as a depopulation of purely metallic junctions), unveiling a rich dynamic response. This procedure triggers the achievement of a capacitive response, which is successfully rationalized using a previously introduced 'two junction model'. Thermal treatment appears to be an indirect strategy to effectively modify the humidity content at the NW-NW intersections and, consequently, enable multiple switching schemes suitable for brain-like processing alternatives.

\end{abstract}

\vspace{0.5 cm}

\noindent Self-assemblies are understood as the autonomous organization of components into patterns without human intervention \cite{whitesides_self-assembly_2002}. %Metallic nanoparticle or nanowire networks, structures formed by the domain boundaries in a ferroic material (either ferromagnetic or multiferroic), percolative paths in infiltrated mesoscopic materials, and/or metal-organic frameworks are just a few examples of self-assemblies. They have been gaining attention due to specific features that resemble biological assemblies with information-processing abilities.
In turn, self-assembled networks formed by metallic nanowires (NWs) are intensively studied as a bio-inspired playground for a neuromorphic alternative to traditional digital computation approaches \cite{markovic_physics_2020,kuncic_neuromorphic_2021}. In this architecture, some features of their biological counterparts are achieved: electrical conductivity \cite{sannicolo_direct_2016}, tunability of internal resistance states \cite{diaz_schneider_resistive_2022}, emergence of collective properties \cite{di_francesco_spatiotemporal_2021}, and a high degree of interconnectivity \cite{chialvo_emergent_2010}. Upon closer examination of the constituents, metallic NWs can be compared to biological axons, while their coating forms capacitor-like structures at the points of intersection between them. The ability to tune the resistance at the cross-point of two coated NWs (NW-NW junction), in a memristive-like way, is comparable to adapting the synaptic weight between two neighboring neurons \cite{zhu_information_2021}. Moreover, the connectivity of the NWs -conditioned by synthesis and electrical measurements \cite{diaz_schneider_two-junction_2024}- determines the intrinsic self-organization of the network, whose conductivity and, consequently, overall resistance are governed by the number of available paths connecting the externally accessible electrodes.

Metallic NWs, especially those made of Ag, are well-known systems since they are suitable for transparent electrodes \cite{langley_metallic_2014,lagrange_optimization_2015,mayousse_stability_2015,guan_performance_2022}. In that context, studies focused on optimizing their stability \cite{lagrange_optimization_2015,guan_performance_2022,bardet_silver_2021}. By then, reducing their sheet resistance and assuring the sustainability of their associated electrical properties were key to boosting their assimilation in the technological flow \cite{mayousse_stability_2015,sharma_review_2022,zhang_recent_2021}. More recently, AgNWNs witnessed a renewed interest due to their wide spectrum of interesting behaviors that aim at being exploited with neuromorphic scopes worldwide \cite{kuncic_neuromorphic_2021,diaz_schneider_two-junction_2024,milano_brain-inspired_2020}. In this case, being able to trigger conductance variations is, to some extent, desirable since it reveals the system's capability to implement a certain type of fading memory \cite{rao_unraveling_2025}. 

In this context, we present a detailed study of the complex impedance response of dense AgNWNs while subjected to thermal and electrical stresses. We have found that a moderate thermal treatment (usually disregarded), combined with the application of an AC signal, has a non-deniable impact on the overall impedance state. By involving previous findings on the environmental influence on the electrical response of these AgNWNs \cite{diaz_schneider_resistive_2022} and considering a previously developed microscopical picture \cite{diaz_schneider_two-junction_2024}, we argue that temperature control becomes a strategy to affect the assembly constituents and modulate their macroscopic state. 

\begin{figure}[ht!]
        \centering
        \includegraphics[width=\columnwidth]{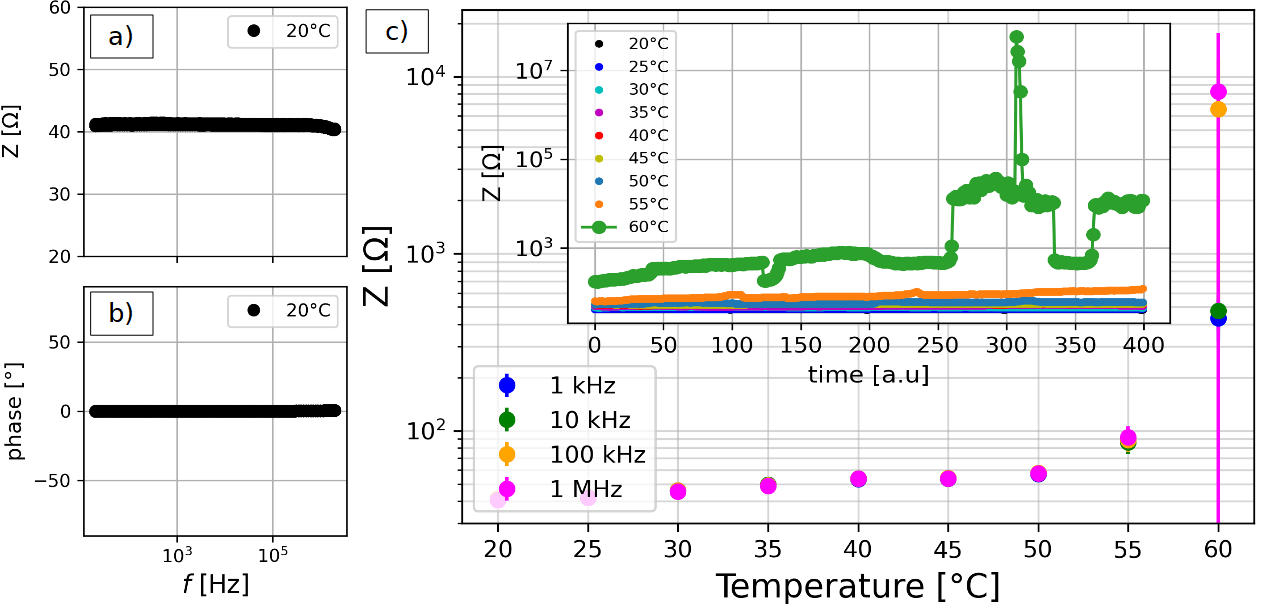}
        \caption{\label{fig:F1} \textbf{Pristine state of dense Ag NWNs.} (a) Impedance’s absolute value, Z, and (b) phase measured as a function of frequency, \textit{f} (from 20 Hz to 2 MHz) with an AC amplitude of 20 mV, a bias voltage set to 0 V, and T = 20°C (RT). (c) Z as a function of temperature, recorded at different \textit{f}. Multiple \textit{f} sweeps from 20 Hz to 2 MHz are successively recorded. Z readings for specific \textit{f} (1 kHz, 10 kHz, 100 kHz, and 1 MHz) were selected and averaged over the four runs. The mean value and standard deviation for each selected \textit{f} are plotted as a function of T as solid circles and error bars, respectively. The inset depicts the time-domain representation of the data, where Z exhibits stochastic, non-monotonic excursions but a gradual overall rise, with no correlation to the excitation frequency.}
        %
        %The inset depicts the same experimental data displayed as a function of time, revealing a Z evolution uncorrelated to \textit{f}.}     
\end{figure}

The AgNWs assemblies were fabricated with a targeted geometry ($\sim$ 170 nm in diameter and $\sim$ 70 $\mu$m in length), then dispersed in a solvent (containing polymeric residues coming from the NWs growth), and finally deposited into a desired substrate to form the networks (AgNWNs) \cite{diaz_schneider_resistive_2022}. The desired areal density ($\sim$ 2500 mm$^{-2}$) is reached by sequential deposition steps \cite{diaz_schneider_resistive_2022}. Two millimetric-sized Ag electrodes (1 mm apart) were sputtered to access the assembly electrically. Mounted inside a Faraday box (to assure electromagnetic isolation), a Zeiss probe station, with a thermal chuck and triaxial micromanipulators, was used to connect the electrodes (avoiding weldings that could imply additional impedance contributions). An ATT$^{TM}$ system was used to control the chuck temperature while an Agilent\footnote{Currently Keysight.} E4980A precision LCR meter$^{TM}$ was programmed using a custom-defined Python routine. Throughout the whole communication, impedance is characterized by applying a 20 mV AC excitation with frequency \textit{f} ranging from 20 Hz to 2 MHz. For each condition, impedance's absolute value, Z, and phase are recorded as functions of \textit{f} (see Figs. \ref{fig:F1} a–b). Standard open/short (and lead) compensation was performed to remove fixture and cable parasitics before analysis. 
%impedance is characterized by applying an AC voltage level of 20 mV with the frequency, \textit{f}, ranging from 20 Hz to 2 MHz. For each condition, Z and phase are recorded as a function of \textit{f} (see Figs. \ref{fig:F1} (a) and (b)). 
The DC bias voltage, V$_{\mathrm{DC}}$, and the external temperature, T, are specified for each experiment.  

\noindent On one hand, we have previously shown that dense assemblies of AgNWs display a pristine low-resistance linear response \cite{diaz_schneider_two-junction_2024}. %By using impedance spectroscopy (Fig. 1) we reinforce that picture by crediting the Ohmic behavior of the dense Ag nanowire network in their pristine condition. 
Although the NWs are coated by a polymer, electrical percolation between the macroscopic electrodes is attained. %Due to this phenomenology, these dense AgNWs are referred to as over-percolated \cite{diaz_schneider_two-junction_2024}. 
%
%We have previously proposed a scenario to rationalize this response: the metallic behavior of the otherwise expected capacitive (eventually memristive) junction between neighboring NWs lies in the existence of pinholes and other defects that likely populate the system, such that a dominant percentage of junctions exhibit this behavior \cite{diaz_schneider_two-junction_2024}.
%
We have previously proposed a scenario to rationalize this response based on the existence of two types of NWs junctions \cite{diaz_schneider_two-junction_2024}. In that picture, the NW-NW junctions could be either capacitive, as expected from a polymer-mediated metal-metal crosspoint, or metallic, due to the formation of conducting filaments that short-circuit the polymeric dielectric \cite{krishnan_mechanism_2016,milano_electrochemical_2024}. In this scenario, although there are plenty of NWs' intersections forming memristive junctions, the metallic ones prevail by predominantly carrying the current flow. %Metallic and memristive junctions compete for the current flow, with the former being more conductive than the latter. 
Impedance measurements (Fig. \ref{fig:F1}) support this scenario by showing a zero-phased contribution (Fig. \ref{fig:F1} (b)) with the absolute impedance, Z, being frequency-independent (Fig. \ref{fig:F1} (a)) and rising as temperature is raised from 20 to 60°C (Fig. \ref{fig:F1} (c)).  
%
%Upon increasing the frequency of the externally applied AC electric field, the impedance associated with the capacitive-like memristive junctions becomes more convenient, and the phase shows hints of moving towards negative values (-90° would represent a pure capacitive response).
%The evolution of the impedance's absolute value (Z) as a function of the external temperature (T) displays a monotonic increase (Fig. \ref{fig:F1} (c)), validating the metallic nature of the system. 
Four \textit{f} sweeps were recorded at each T. %Further details are included in Figure S1 of the Supplementary Information (SI), where all the programmed and measured values are unfolded as a function of time. 
It is thus possible to extract the mean value and standard deviation for different frequency conditions (Fig. \ref{fig:F1} (c)). No difference among the four different selected frequencies (1 kHz, 10 kHz, 100 kHz, 1 MHz) was observed at each T while T $<$ 55°C, as expected from the metallic behavior. Noticeably, the pronounced increase towards T $\sim$ 60°C and the obtained dispersion suggest the onset of a distinct regime. %While at middle T there are no appreciable differences between multiple frequency conditions (reflection of the purely resistive nature of the sample), from 55 °C on, those values significantly differ from each other. Figure S1 displays the data plotted as a function of time (while sweeping frequency), demonstrating the onset of a regime change at $\sim$ 60°C with Z varying across multiple orders of magnitude. 
The inset in Fig. \ref{fig:F1} (c) contains the same experimental data displayed as a function of time. This graph helps in understanding the magnitude of the pronounced error bars observed in Fig. \ref{fig:F1} (c) at 60°C. While low dispersion was recorded for T < 55°C, the values obtained at 60°C span the entire impedance range. Figure S1 of the Supplementary Information (SI) includes the programmed (\textit{f}) and measured (Z and phase) quantities unfolded as a function of time. Z evolves apparently uncorrelated to \textit{f}. The phase lies close to zero during the complete measurement (see Fig. S1), depicting subtle variations for high frequencies: bending towards positive values for moderate Z ($<$ 10$^3 ~ \Omega$), becoming negative for higher Z ($\sim$ 10$^5 ~ \Omega$). Moreover, the instability, triggered by the combination of temperature and a repeated electrical stimulating strategy, reveals a completely different phenomenology afterward.

\begin{figure}[ht!]
        \centering
        \includegraphics[width=0.7\columnwidth]{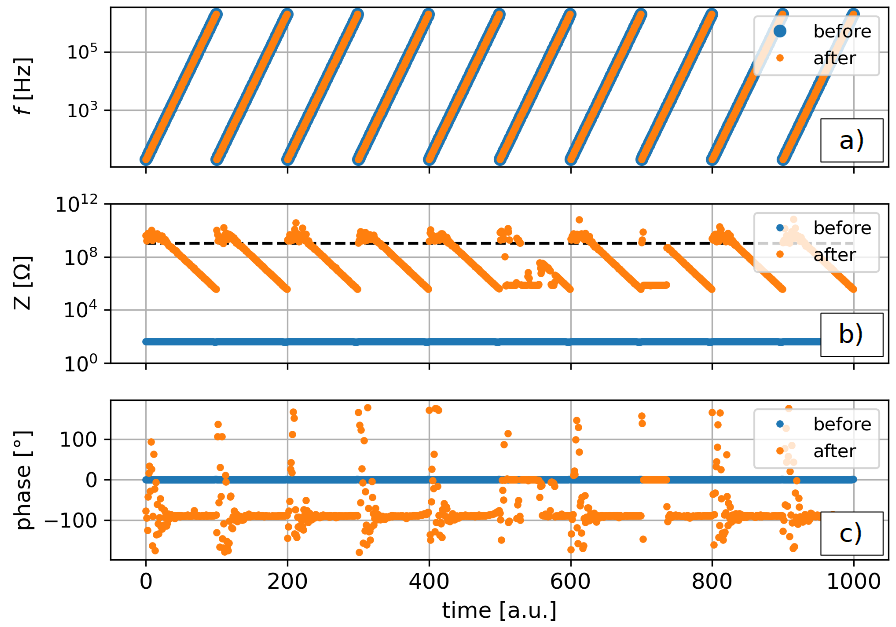}
        \caption{\label{fig:F2} \textbf{Before and after low-T annealing impedance responses recorded at RT.} (a) Ten frequency, \textit{f}, sweeps between 20 Hz and 2 MHz (AC amplitude = 20 mV and V$_{\mathrm{DC}}$ = 0 V) were applied to record the complex impedance response. The dependencies of (b) Z (logarithmic scale) and (c) phase (linear scale) as a function of \textit{f} illustrate the contrast between the ohmic response in the pristine condition and the capacitive dependence obtained before and after the conducted annealing, respectively. In addition, Z exhibits intermittent instabilities, abruptly switching between capacitive and purely resistive behavior, consistent with sudden transitions from a high-resistance state to intermediate resistance levels. The black dashed line in Z indicates the instrumental limit (1 G$\Omega$ \cite{lcr-meter}).} % Ten successive runs of complex impedance measurements were performed at RT before and after the temperature excursion up to 60°C. (d) By fitting the dependence of Z as a function of f, the equivalent capacitance is determined. Under a few assumptions, the capacitance of each junction is obtained and, in turn, related to the expected polymer dielectric permittivity, which is here (d) plotted as a function of the geometrical factor of the nanowires' assembly (see text).}     
\end{figure}

After cooling down to room temperature, RT $\sim$ 20°C, the impedance measurement is repeated. Ten \textit{f} sweeps were implemented and compared (same experiment performed immediately before ramping up T). Figure \ref{fig:F2} presents the \textit{f}, Z, and phase displayed as a function of time. The routines were strictly the same (as indicated by Fig. \ref{fig:F2} (a) with V$_{\mathrm{DC}}$ = 0 V) and T $\sim$ 20°C. However, the electrical response was completely different. The dependence of Z on a logarithmic scale and the phase value indicate that the nature of the sample after the annealing is capacitive (with a clear reduction of Z as a function of the frequency increase and a phase locked to -90°). The impedance state is so dramatically different that low-frequency impedance values are hardly measured due to the instrumental limits (1 G$\Omega$ \cite{lcr-meter}, indicated as a black dashed line in Fig. \ref{fig:F2} (b)). This, in turn, allows explaining the instability in the phase close to that condition. Furthermore, Z presents instabilities characterized by abrupt switches between capacitive and purely resistive responses, in line with the jumps from a high-resistance state to intermediate resistance levels.

\noindent The dramatic change in the impedance state, produced during the T reduction, is remarkable, considering the gentle annealing applied to the system (20 to 60°C). Much higher annealing conditions have been applied to systems like this without the appreciation of any clear structural change \cite{lagrange_optimization_2015,guan_performance_2022,bellet_transparent_2017}. Still, this behavior was observed in other samples (Fig. S2, Supplementary Information). The solution to this dilemma probably lies in our previous observation regarding the environmental impact on the electrical response \cite{diaz_schneider_resistive_2022}. Raising T not only implies heating the sample, but also favors the release of water molecules adsorbed at the junctions. The water content present at the polymeric junction between the metallic nanowires enables the formation of conducting filaments due to the enhanced mobility of the Ag$^+$ ions. In turn, the formation or dissolution of the conductive filament determines the metallic or capacitive behavior of the NW-NW junctions, in tandem with an increase to a high-resistance regime. Thus, water availability, coming from the humidity of the environment, would provide the means that promote short-cutting the otherwise expected capacitive stack at each intersection. Increasing T indirectly favors the water desorption, potentially promoting dissolution of the metallic filament \cite{mayousse_stability_2015,grillet_photo-oxidation_2013,jiu_effect_2015} and, consequently, the recovery of a capacitive behavior of the junctions. Water removal from the junctions could indirectly impact the filament dissolution either by accelerating oxidation of the Ag filament in air or due to the heating provoked by the AC stimulus, which cannot be dissipated through the water (with a high caloric capacity), consequently burning the ionic filament. Considering the evidence in the literature \cite{mayousse_stability_2015,grillet_photo-oxidation_2013,jiu_effect_2015}, the second explanation is more likely since the combination of water and temperature is reported to promote oxidation and electrical collapse. Thus, the water removal would be indirectly responsible for the extreme resistance variation reported in our previous communication \cite{diaz_schneider_resistive_2022}, close to laboratory conditions, in which relative humidity may vary in a wide range. Figure S2 presents the live variation of the resistance level on the timescale of minutes recorded as a function of time under a constant voltage stress (100 mV). The huge variation seems to eventually get exhausted (variation is observed only in the first run after thermal stressing), at least until higher  V$_{\mathrm{DC}}$ levels are applied. 
%
%The dramatic change in the impedance state recalls the electrofusing operation required for dense AgNWN under strict DC schemes \cite{diaz_schneider_two-junction_2024}. Low-impedance pristine AgNWN were previously subjected to an electro-thermal operation referred to as electrofusing, which would erase the more conducting paths of the assembly and reveal a more resistive and broader variety of electrical responses. The experiment conducted here (combining AC signal and thermal stimulus) supports the idea that electrofusing is just an experimental strategy (among others) to edit the assembly connectivity. Electrical percolation is attained considering the network geometry in addition to the presence of metallic junctions (in pristine highly-dense AgNWNs, 96 $\%$ of the total of junctions are metallic \cite{diaz_schneider_two-junction_2024}). Here, we argue that temperature modulation affects the population of metallic junctions, and concomitantly, the number of 'effective' paths through which the current flows. This effect is equivalent to a shift in the percolation curve towards higher NWs' densities to achieve a low resistance state. 

Previously, the way to overcome the highly conductive pristine state and access other conductive states was through “electrofusing” (EF) \cite{diaz_schneider_two-junction_2024}, a procedure in which the paths with the highest conductivity are burned after applying an excessive current. Otherwise, the persistence of high-conductance connections overshadows the presence of more tunable filament-mediated memristive junctions. Here, implementing a combination of AC stimuli and involving a moderate thermal budget, the environmental influence is utilized in favor of the desired phenomenology. This strategy is, at the same time, an alternative to EF, since it unlocks a rich variety of impedance states (avoiding a less convenient electrical procedure), and a complement, because the thermal treatment is homogeneous (while the EF is strongly inhomogeneous). The effectiveness of the thermal procedure supports the idea that electrofusing is just an experimental strategy (among others) to edit the assembly connectivity. Electrical percolation is attained considering the network geometry in addition to the presence of metallic junctions (in pristine highly-dense AgNWNs, 96 $\%$ of the total of junctions are metallic \cite{diaz_schneider_two-junction_2024}). Here, we argue that temperature modulation affects the population of metallic junctions, and concomitantly, the number of 'effective' paths through which the current flows. This effect is equivalent to a shift in the percolation curve towards higher NWs densities to achieve a low resistance state. Being the percolation curve dependent on the percentage of the population of metallic junctions, this quantity could be thought of as a third dimension in the percolation dependence. 
 
The capacitive state revealed after the thermal excursion is suitable for parameter extraction. By recalling that the impedance of a pure capacitor is described as $Z = \frac{1}{2 \pi f \ C}$, linearly fitting the double logarithmic plot of log$_{10}$Z as a function of log$_{10}$\textit{f} (not shown but similar to Figure S3) allows to extract a value of C$_{eq}= 0.3$ pF. Whether this value comes from the AgNWN or from an undesired, poorly compensated setup contribution is debatable. Wires' contribution could be of the order of pF \cite{lcr-meter}. However, comparison among multiple measurements of the same sample and different samples (see Figure S3) sheds light in this regard. Differences are found in the capacitance recorded for each case, indicating that the specific condition of the sample is responsible for (at least) part of the complex response. Values of C$_{eq}= 0.06-10$ pF are obtained through all the samples and conditions measured during this study. Being the capacitive term (C$_{eq}$), the result of measuring the whole AgNWN, i.e., the overall response of the self-assembly using the two macroscopic electrodes and the instrumental connections, rationalizing the estimation in terms of the nanoscopic components, although speculative, is possible. The high density of nanowires per unit area ($\sim$ 2500 mm$^{-2}$) suggests an underlying equivalent circuit comprising multiple parallel and serially connected conducting paths. Following this reasoning and assuming a geometrical arrangement of the NWs (see Supplementary Information), the individual contribution of each memristive junction could be estimated. However, the indirectly extracted values are hard to rationalize (see Figure S4), probably due to the extremely naïve involved assumptions. For that reason, the effective capacitance of the whole network is the only representative value to be considered in the following. Moreover, the obtained capacitive term fairly agrees with the term obtained in a recent impedance study performed with Ag nanoparticles \cite{rao_unraveling_2025}. %Although in that case the equivalent circuit comprises a constant phase element, instead of a pure capacitor, the contribution lies in the order of the pF with physical dimensions similar to those involved in the NW networks presented here.   

\begin{figure}[ht!]
        \centering
        \includegraphics[width=0.7\columnwidth]{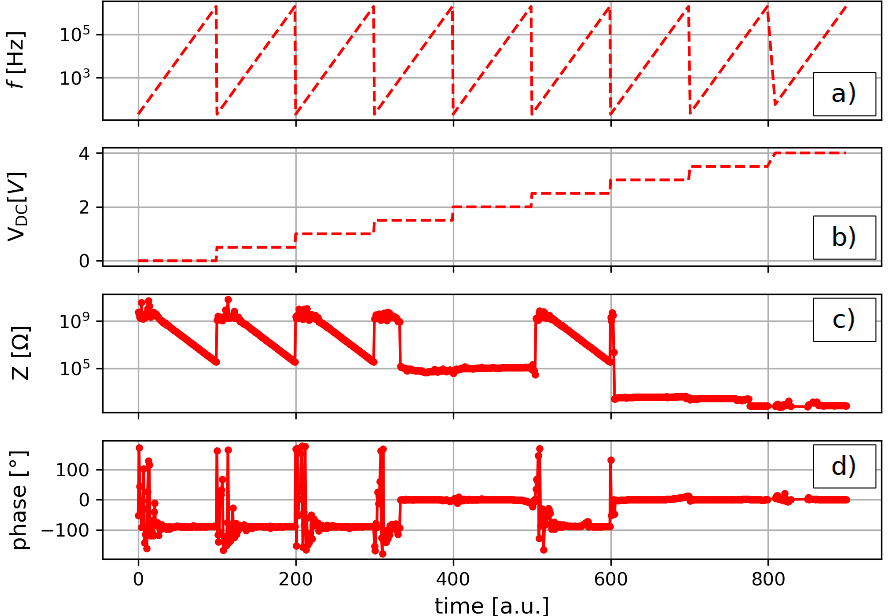}
        \caption{\label{fig:F3} \textbf{Successive runs of complex impedance measured upon different DC bias voltage conditions.} (a) \textit{f} and (b) V$_{\mathrm{DC}}$ programmed for the sequential runs. (c) Z and (d) phase recorded as a function of time. All the measurements were conducted using an AC amplitude of 20 mV.}
\end{figure}

\noindent Electrically stressing the sample, either by pure AC stimulus (Figure \ref{fig:F2}) or combined AC-DC stimuli (see Figure \ref{fig:F3}), the sample returns to a stable low-impedance state. Figure \ref{fig:F3} illustrates how applying increasing V$_{\mathrm{DC}}$ values, the low-impedance purely-resistive regime can be recovered (without intentionally varying temperature). A sequential increase of the DC bias while recording the impedance indicates the competition between a capacitive-like dependence and the restoration of a purely resistive nature. Close to 1 V, a purely resistive (Z $\sim$ 10$^3~\Omega$) zero-phased state is achieved. Further increase of the bias voltage provokes the return of the AgNWN to a capacitive condition, most likely indicating a competition between the DC stimulus, which favors the filament formation, and the AC signal, which promotes them to vanish. Eventually, the DC driving force is so intense that a newly purely resistive state (Z $\sim$ 10$^1~\Omega$) is achieved and persists for a longer period. %This illustrates the interplay between thermal and electrical stimuli, the former apparently favoring filament dissolution while the later would restore them. 

Figures S5 and S6 help clarify the situation. Figure S5 illustrates the interplay between the AC and DC stimuli, with moderate DC values ($\sim$ 1 V) producing a short-term impedance change and higher DC values (beyond 2 V) triggering a more sustainable final state. Moreover, a careful measurement performed with a smaller DC bias step (Figure S6) depicts a previously recorded phenomenology. Ramping up the bias voltage starts to demonstrate switching towards higher impedance (still metallic) states. Close to 0.3 V, there is an onset of a switching regime from low to higher resistance (50 $\Omega$ to 100 $\Omega$, in this case) which is in agreement with previously observed phenomenology in percolated and over-percolated AgNWNs (sparse to dense assemblies, respectively) \cite{diaz_schneider_two-junction_2024}. %One way to access other conductive states and observe resistive switching is through “electrofusing” \cite{diaz_schneider_two-junction_2024}, a procedure in which the paths with the highest conductivity are burned after applying an excessive current. Otherwise, the persistence of high-conductance connections overshadows the presence of more tunable filament-mediated memristive junctions. Here, implementing a combination of AC-DC stimuli and involving a moderate thermal budget, the environmental influence is utilized in favor of the desired phenomenology, helping to avoid a less convenient electrical procedure (electrofusing). 

\begin{figure}[ht!]
        \centering
        \includegraphics[width=0.7\columnwidth]{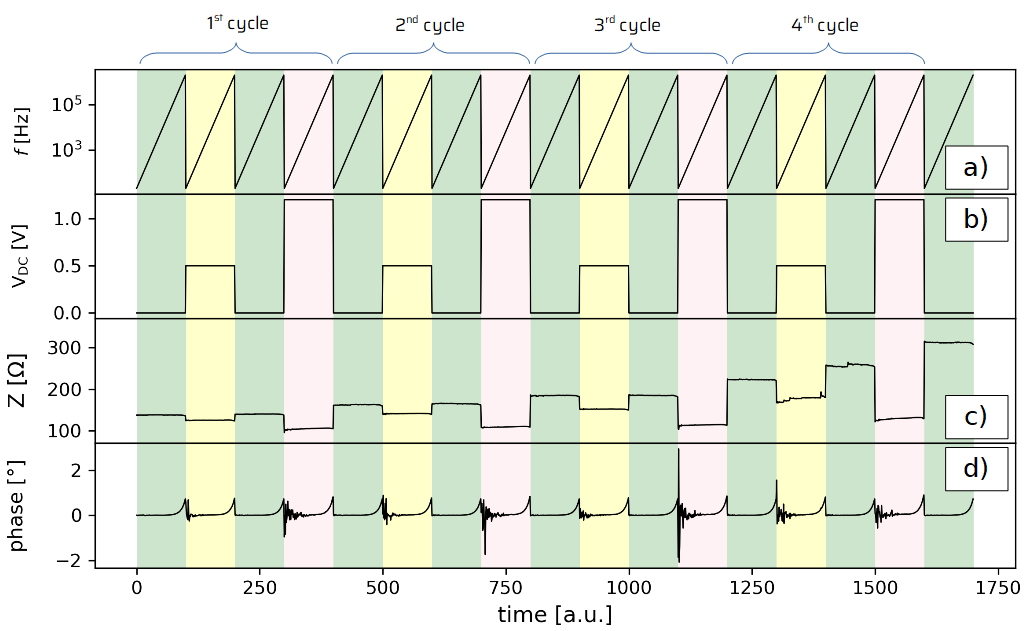}
        \caption{\label{fig:F4} \textbf{Bias voltage sequences.} Successive runs of complex impedance measured as a function of the \textit{f} (from 20 Hz to 2 MHz) recorded at different bias voltages. Four cycles of V$_{\mathrm{DC}} = \{0;0.5;0;1.2\}$ V are implemented to observe the progression of Z and phase.}
\end{figure}

Finally, having argued that impedance unveils two different switching schemes (short and long term) depending on the DC bias voltage, implementing a sequence of different values provides new insights. Figure \ref{fig:F4} represents a stepped sequence that sets V$_{\mathrm{DC}} = \{0;0.5;0;1.2\}$ V repeatedly. The two non-zero values are aimed at provoking short- and long-term changes, respectively, while the remnant readings determine whether the achieved states are persistent or not. Four cycles of the V$_{\mathrm{DC}} = \{0;0.5;0;1.2\}$ V sequence were programmed (Figure \ref{fig:F4}). The first and second sequences demonstrate that 0.5 V does not impact the remnant state measured immediately afterward, while the 1.2 V stimulus produces a persistent effect, revealed as an impedance increase at the remnant condition. During the third sequence, either due to the accumulation of applied stimuli or to the impedance increase (or both), V$_{\mathrm{DC}} = 0.5$ V does impact the persistent state of the network. %Further work is required to distinguish between these two explanations (based on the cumulative effect and/or the impedance level). 
%Moreover, the impedance reduction during the application of the non-zero bias values is counterintuitive considering the previously presented results (Fig. \ref{fig:F3}). Whether this is the consequence of further evolution of the system upon environmental influence (producing temporary changes not related to the particular applied stimulus), or a sort of electrostatic/ferroelectric effect \cite{lovinger_ferroelectric_1983}, is still unclear. 
At the same time, carefully analyzing the impedance phase, in all the sweeps programmed in this experiment, there is a tendency towards positive values of the phase, suggesting an inductive contribution. Whether this comes from the experimental setup, and it gets competitive whenever the absolute value becomes higher, or the behavior strictly relates to the nanowire network itself, is discussed in the Supplementary Information. %Repeating the experiment but programming 10 times more points per frequency decade gives the same results (see Figure S7 in the Supplementary Information): a subtle phase increase towards the end of the frequency range (close to 1 MHz) and a maximum value close to 1°. Being the inductive term so small makes it difficult to fit and, consequently, to estimate the inductive contribution responsible for such an effect.    

In summary, the described phenomenology, supported by impedance measurements, complements the previously introduced quasi-DC electrical characterization, opening the way to neuromorphic applications beyond synaptic weight storage. Using the environmental impact on the electrical response, we have been able to unlock a new region of the percolation diagram, considering the $\%$ of metallic junctions as its third dimension. Thermal tuning of the system, assisted by AC signal stimuli, is a successful strategy that utilizes the extreme sensitivity of these samples to our advantage. At the same time, rationalizing the results obtained by using the predictions of our previously introduced model reinforces its validity. Finally, being able to modify the overall response with this method is considerably more convenient than electrofusing, the latter being an electrical method capable of disconnecting highly conductive paths by breaking wires, albeit at the expense of high power dissipation. The combination of frequency exploration and thermal stimulus showcases multiple switching regimes on different timescales while allowing for the identification of instabilities suitable for spike production and in-hardware soma implementation.

\section*{Acknowledgements}

The authors kindly acknowledge Prof. Juan Claudio Nino and the Ph.D. candidate Chaitanya Sharma from the University of Florida (USA) for their insightful comments and the fruitful discussions. The authors are also grateful to Dr. Mariano García-Inza (CONICET - FRBA/UTN - FIUBA, Argentina) for technical assistance and the Electronic Department National Technical University (regional faculty of Buenos Aires, Argentina) for giving us access to conduct the temperature-controlled impedance measurements.

This work was supported by CONICET PIP 2023-2025 11220220100508CO and ANPCyT-FONCyT PICT 2021-0876. EDM acknowledges financial support from CONICET PIBAA 2022-2023 28720210100473CO. CPQ acknowledges financial support from CONICET PIBAA 2022-2023 28720210100975CO. JIDS acknowledges a fellowship from CONICET.

This research has also benefited from the exchanges supported by the Maldacena project (from the Balseiro Institute, 2024 and 2025 editions), and the Fulbright program (2024).

\printbibliography

\begin{center}
    \huge{Thermal tuning of dynamic response in Ag-based nanowire networks}\\\textbf{SUPPLEMENTARY INFORMATION}
    \\\large{J.I. Díaz Schneider, C. Gómez, C. Acha, P.E. Levy, E.D. Martínez, C.P. Quinteros}
\end{center}

\section*{Impedance as a function of temperature and its temporal evolution}

Displaying the complex impedance measurements as a function of time (while still sweeping the frequency), recorded at different temperatures, it becomes clear that the evolution depends strongly on the thermal budget. The drift in the mean Z value recorded at 55 and 60°C is much more noticeable than for the previously measured conditions. The variability is so marked that the Z value covers almost the full range of measurable values (from 10$^{1}$ to 10$^{8}$ $\Omega$). This evolution justifies the increase in the standard deviation for the highest T values presented in Figure 1 of the Main. Changes are not related monotonically to the ramping frequency, thus allowing us to consider that the instabilities are independent of the applied stimulus.    

\begin{figure}[ht!]
        \centering
        \includegraphics[width=0.7\columnwidth]{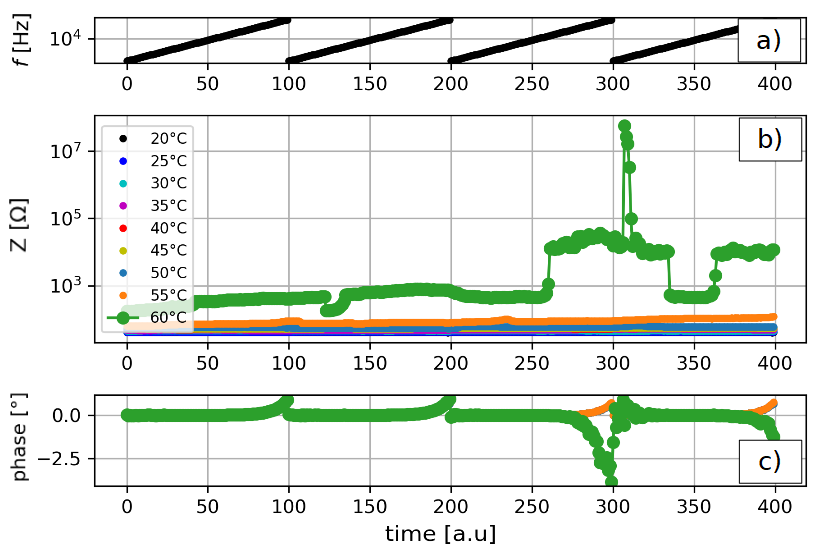}
        \caption{\label{fig:FS1} \textbf{Temporal evolution of the impedance depending on the external temperature}. (a) Frequency, \textit{f} (b) impedance's absolute value, Z, and (c) phase as a function of time, for different  external temperatures.}     
\end{figure}

\section*{Variability in quasi-DC resistance measurement as a function of time}

After thermally cycling a AgNWN, a quasi-DC resistance was recorded while holding 100 mV applied between the two macroscopic electrodes. The results are presented in Figure S2. Dramatic changes are observed during the development of the measurements themselves (of the order of minutes). This measurement, recorded in a different AgNWN than the one presented in Figures S1, 1, and 2, credits that the same spontaneous fluctuation is obtained in multiple samples and also that variation could occur regardless of the AC applied stimulus. % spontaneous variations can be as high as observed between the successive runs presented in Figures 2 and 3 of the Main. 

\begin{figure}[ht!]
        \centering
        \includegraphics[width=0.7\columnwidth]{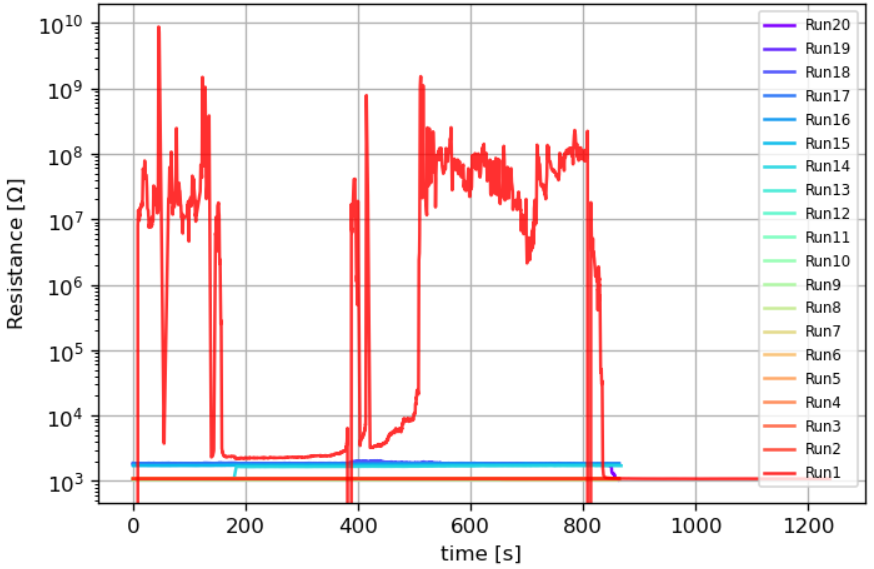}
        \caption{\label{fig:FS2} \textbf{Spontaneous evolution of the quasi-static resistance at room temperature}. After thermal cycling, a constant voltage is applied (100 mV) while recording the current as a function of time. Defined as the ratio between voltage and current, the resistance is here displayed as a function of time. The first Run demonstrates a dramatic variation (covering the whole experimental resistance range). Further runs show an intermediate resistance value ($\sim$ 2 k$\Omega$) with smaller variations, suggesting the instabilities could represent a transient response.}     
\end{figure}

%Fig. S2 (Supplementary Information) displays the results of a constant stress voltage experiment (DC voltage = 100 mV). Calculated as the ratio between voltage and current, the quasi-DC resistance shows dramatic changes as a function of time. The sample had been previously subjected to thermal and electrical stimulation as the sample presented in Figs. 1 and 2. 

\newpage
\section*{Rationalizing capacitive contribution}

\begin{figure}[ht!]
        \centering
        \includegraphics[width=\columnwidth]{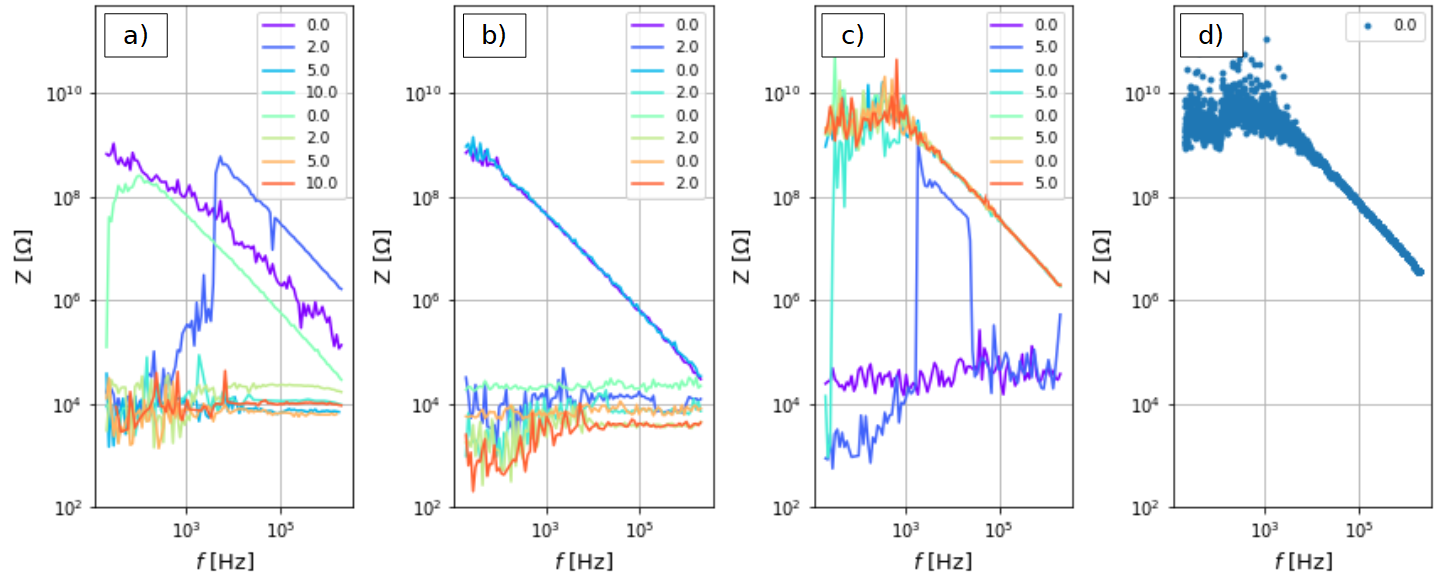}
        \caption{\label{fig:FS3} \textbf{Absolute impedance recorded in different samples and multiple runs on the same sample under different biasing conditions.} Differences among the presented measurements, recorded in the same experimental setup, illustrate capacitive (linear with -1 slope) and resistive (constant) dependencies as a function of \textit{f}. (a), (b), and (c) Z recorded as a function of \textit{f} for the same sample measured under different DC bias voltages. (d) Z-\textit{f} response of an AgNWN with an enlarged NW-NW junction thickness. }     
\end{figure}

Wires, adapters, micromanipulators, sample contacts, and even the ports of the measuring instrument (LCR meter \cite{lcr-meter}) contribute to the total impedance. Even though the LCR meter reduces the influence of the experimental setup (by properly conducting the open, short, and load compensation \cite{lcr-meter}), it always persists as an undesired contribution. In this situation, a capacitance of the order of pF could be perfectly attributed to the uncompensated setup. However, Figure S\ref{fig:FS3} reveals additional information. 

\noindent Figure S\ref{fig:FS3} (a), (b), and (c) display the Z response upon sweeping \textit{f} (Z-\textit{f}) under different DC bias voltages (correspondingly labeled in each case). Depending on the DC bias and on the history of applied stimuli, the sample depicts either a capacitive (linear with -1 slope) and resistive (constant) Z-\textit{f} dependence. For those measurements admitting a linear fit with a negative slope, the equivalent capacitance is determined from the intersection with the y-axis. Different values of equivalent capacitance (C$_{eq}$), obtained from the same sample under multiple DC bias conditions (labeled in each panel), indicate that this attribute cannot be attributed solely to uncompensated artifacts. There must be a dependence on the sample state that explains variations in successive runs. Complementarily, Figure S\ref{fig:FS3} (d) shows the dependence recorded in a different AgNWN (with thicker NW-NW junctions) measured in the same experimental setup. Thus, we argue that although the presence of a spurious contribution due to the wires cannot be ruled out, this variation suggests that part of the capacitive response originates solely from the samples themselves.     

\subsection*{Estimation of individual junctions' capacitance using a geometrical model}

Different equivalent capacitance values (C$_{eq}$) were obtained for multiple samples recorded under different polarization conditions. Obtained C$_{eq}$ values range from 0.6 pF to 10 pF. Considering the NWs assemblies as a collection of NW-NW junctions with a capacitive nature, here we aim at relating the indirectly obtained C$_{eq}$ values with the underlying characteristics of the assemblies.

Assuming M parallel paths composed of N serially connected identical capacitors would give an expression to relate the equivalent capacitance to the individual value of each junction C$_{eq} = \frac{M}{N} \cdot$ C$_i$. Depending on the number of paths for the current to flow, C$_i$ could be extracted. %Nevertheless, the assumption that each junction possesses a fixed capacitance is most likely an oversimplified picture of the situation. 

\begin{figure}[ht!]
        \centering
        \includegraphics[width=0.7\columnwidth]{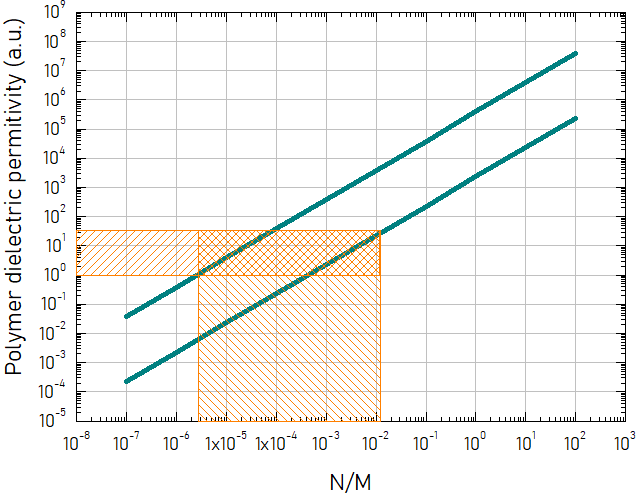}
        \caption{\label{fig:FS4} \textbf{Relationship between polymer dielectric permittivity and the geometrical factor} (corresponding to the ratio between serial and parallel connected capacitors). The two lines correspond to the relationship deduced from using the two extreme obtained capacitances (C$_{eq}$ = 0.6 pF and 10 pF). A reasonable permittivity for a polymer would lie within 10$^0$-10$^1$, which would correspond to 10$^{-6} < \frac{N}{M} < $ 10$^{-2}$.}     
\end{figure}

\noindent To relate the physical characteristics of the NW-NW junctions with the individual contribution C$_i$, a parallel plate model could be invoked (C$_i = \epsilon_0 \cdot \epsilon \cdot \frac{area}{thickness}$). Assuming that each C$_i$ contribution corresponds to an intersection between two NWs of the network, its associated $area$ and $thickness$ would be: the NWs' diameter to the square $\sim$ (170 nm)$^2$ \cite{diaz_schneider_two-junction_2024}, and double of the coating width ($\sim$ 10 nm \cite{diaz_schneider_resistive_2022}), respectively. By conveniently replacing the estimated values (C$_i = \epsilon_0 \cdot \epsilon \cdot \frac{(170~nm)^2}{10~nm}$), a toy model could be used to relate the relative permittivity ($\epsilon$) and the geometrical factor ($\frac{N}{M}$), see eq. \ref{eq:Ci}. 

\begin{equation}
    C_{eq} = \frac{M}{N} \cdot C_i = \frac{M}{N} \cdot \epsilon_0 \cdot \epsilon \cdot \frac{area}{thickness} \rightarrow \epsilon = C_{eq} \cdot \frac{N}{M} \cdot \frac{thickness}{\epsilon_0 \cdot area}
    \label{eq:Ci}
\end{equation}

\noindent By recalling that the dielectric at each junction would be the polymer remaining from the synthesis, we could inversely extrapolate which would be the geometrical factor ($\frac{N}{M}$) associated with the expected values of $\epsilon$ for polymers (Figure S\ref{fig:FS4}). %Scientific literature reports values close to $10^1-10^2$ for polymers \cite{}. This range of values would correspond to $\frac{N}{M} \sim$ 10$^{-5}$-10$^{-4}$.
%
%Figure S3 presents the polymer dielectric value requested to satisfy the previously described equations as a function of the geometry factor. 
Usual polymer dielectric values range from unity to tens (at most) \cite{tan_advanced_2006}, which would set the geometry factor in the range between 10$^{-6}$ and 10$^{-2}$. %Even in the simplest picture, this estimation would be difficult to rationalize, indicating it is extremely simplistic to relate to the internal arrangement of individual capacitive and resistive contributions.        
This would imply the system would have hundreds to millions more parallel paths than serially connected ones (M $\sim$ 10$^{2-6}$ N). The range of possible $\frac{N}{M}$ values, indirectly derived from these measurements, appears to be so broad that it would be hard to conclude anything based on it. 

There is an alternative way to extract the geometrical factor ($\frac{N}{M}$). Recalling that NWs' areal density $d_a$ corresponds to $\sim 2500~mm^{-2}$, we could consider $N \cdot M$ as the amount of NWs in the active area $n_{NWs} = 2500$. The AgNWN fills the space between two macroscopic electrodes, 1 mm apart and approximately laterally restricted to 1 mm, giving an active area of 1 mm$^2$. Assuming the mean value for the length and width of the NWs, fewer than 15 NWs would suffice to cover the distance between electrodes ($\frac{1 mm}{70 \mu m}$). This reasoning allows obtaining $M = \frac{2500}{15} \sim 170$. Thus, $\frac{N}{M} = \frac{15}{170} \sim 10^{-1}$. This differs from the previous calculation. Moreover, N $\sim 15$ was estimated considering a highly anisotropic sample: NWs' axes almost perfectly aligned with the inter-electrode separation. A more isotropic model would increase the value of N, accompanied by a concomitant reduction in M, and an even stronger underestimation of the geometrical factor. 

This reasoning shows that the physical interpretation of the obtained capacitance is difficult, most likely, due to the mixed contribution of 1. undesired effects from the setup and 2. the AgNWNs themselves. The inability to disentangle the two and the naïve assumptions to electrically model the assembly most likely justify the lack of physical meaning observed. 

%The range of possible $\frac{N}{M}$ values, indirectly derived from these measurements, appears to be overestimated, likely due to a naïve assumption that the NWs are highly anisotropically arranged and perfectly aligned with the direction of the electrical contacts.

\section*{DC-AC impedance variation}

\begin{figure}[ht!]
        \centering
        \includegraphics[width=0.6\columnwidth]{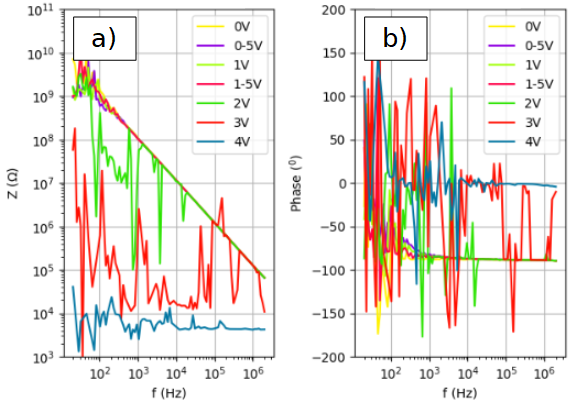}
        \caption{\label{fig:FS6} \textbf{Impedance measurements recorded at multiple DC bias voltage conditions (up to 4 V).} Consistent with Fig. S5, another AgNWN demonstrates the same phenomenology. A capacitive dependence of (a) Z and (b) phase as a function of \textit{f} is recorded for moderate bias voltages ($<2$V). Upon increasing them, the system starts switching to an ohmic behavior characterized by a constant Z and zero phase.}     
\end{figure}

Different samples were subjected to multiple DC bias voltages while recording the complex impedance (AC stimulus required). By conveniently ramping the DC bias voltage, and depending on the initial state, high-to-low or low-to-high impedance switching events are obtained. For an initially high-impedance capacitance state (Fig. S\ref{fig:FS6}), a progressive increase of the DC bias up to 4 V results in a final purely resistive low-impedance state. Unstable states are obtained for the intermediate bias conditions. Sufficiently high DC bias voltages (such as units of Volts and beyond) promote filament formation at the junctions, enabling additional paths for the current flow. Additionally, the rearrangement of the voltage drop within the network, as well as the increased Joule dissipation due to the dramatic impedance reduction, affects the overall impedance and reverts the achieved state. This is observed in Figures 3 and S\ref{fig:FS6}, where capacitive high-impedance dependencies are recovered when the DC voltage is increased. 

\begin{figure}[ht!]
        \centering
        \includegraphics[width=0.6\columnwidth]{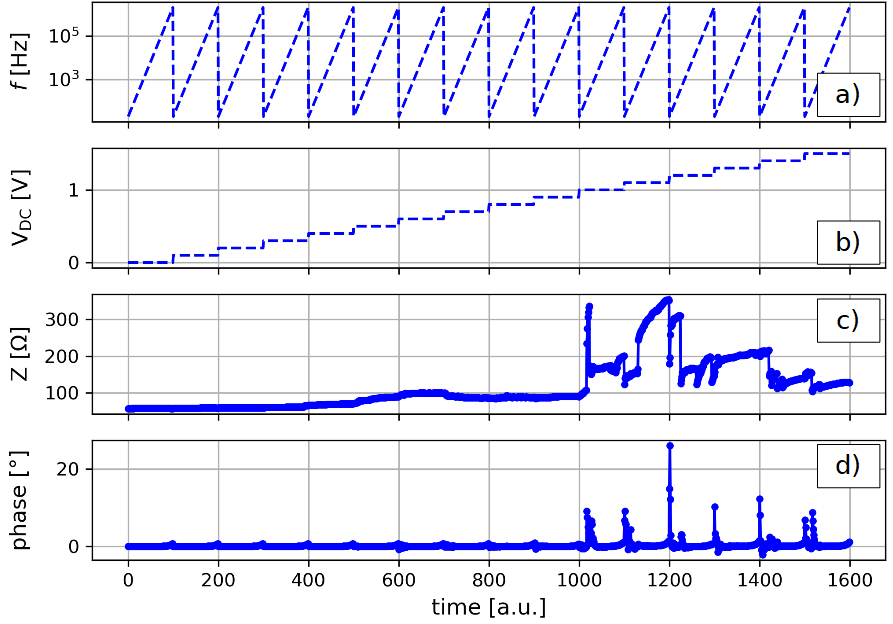}
        \caption{\label{fig:FS5} \textbf{Impedance measurements as a function of moderately increasing DC bias voltage ($<$ 1.5 V).} (a) The frequency is swept during each (b) DC bias voltage applied and depicted as a function of time. (c) Z and (d) phase are displayed as a function of time.}     
\end{figure}

On the contrary, an initial purely-resistive low-impedance state ((Fig. S\ref{fig:FS5})), becomes progressively higher impedance while retaining its purely resistive nature. This behavior is more comparable to the phenomenology described in a previous work \cite{diaz_schneider_two-junction_2024}, where switching from low to high resistive states (both linear in the current vs voltage diagram) was reported. The DC voltage threshold observed here (V $\sim$ 0.3 V) appears smaller than the values previously reported (V $\sim$ 1 V). However, additional measurements demonstrate dispersion (not shown) in this threshold. Moreover, the AC stimulus simultaneously applied in the experiments performed in the current study embodies additional energy that could impact the DC intensity required to observe the referred change.  

%Sufficiently high DC bias voltages (such as units of Volts and beyond) promote filament formation at the junctions, enabling additional paths for the current flow. Additionally, the rearrangement of the voltage drop within the network as well as the increased Joule dissipation due to the dramatic impedance reduction, affects the overall impedance and reverts the achieved state. This is observed in Figures 3 and S\ref{fig:FS6}, where capacitive high-impedance dependencies are recovered when the DC voltage is being increased. %Competition between the DC bias voltage and the elapsed time of the applied stimuli (which relates to the dissipated power) is reflected as contradictory reactions of the impedance level to multiple DC bias conditions. This could explain the difference in the obtained phenomenology in Figures 3 and 4 of the Main.   

% \begin{figure}[ht!]
%         \centering
%         \includegraphics[width=0.6\columnwidth]{Figures/FigureS6.PNG}
%         \caption{\label{fig:FS6} }     
% \end{figure}

\newpage
\section*{Inductive contribution at high frequencies}

Looking closer at the phase for the low impedance states, there is an increase towards high frequencies. Although the increase of the phase is moderate, it appears repetitively for the highest values of the AC frequency. The vertical scale in Figure 3 (of the Main) emphasizes this contribution. Whether this term arises from the AgNWN itself or from the wires is unclear. Disentangling whether this effect originates from the frequency value or from the elapsed time of the whole measurement could shed light on its origin. If this feature were to proceed from the attributes of the assembly, it would most likely be dependent on the specific frequency condition, regardless of the time it takes to conduct the complete measurement. If it were due to some inductive contribution within the wires of the experimental setup, it would be more dependent on the total time of stimulation. However, by programming frequency sweeps of different durations, we could disentangle the frequency and the   

\begin{figure}[ht!]
        \centering
        \includegraphics[width=0.6\columnwidth]{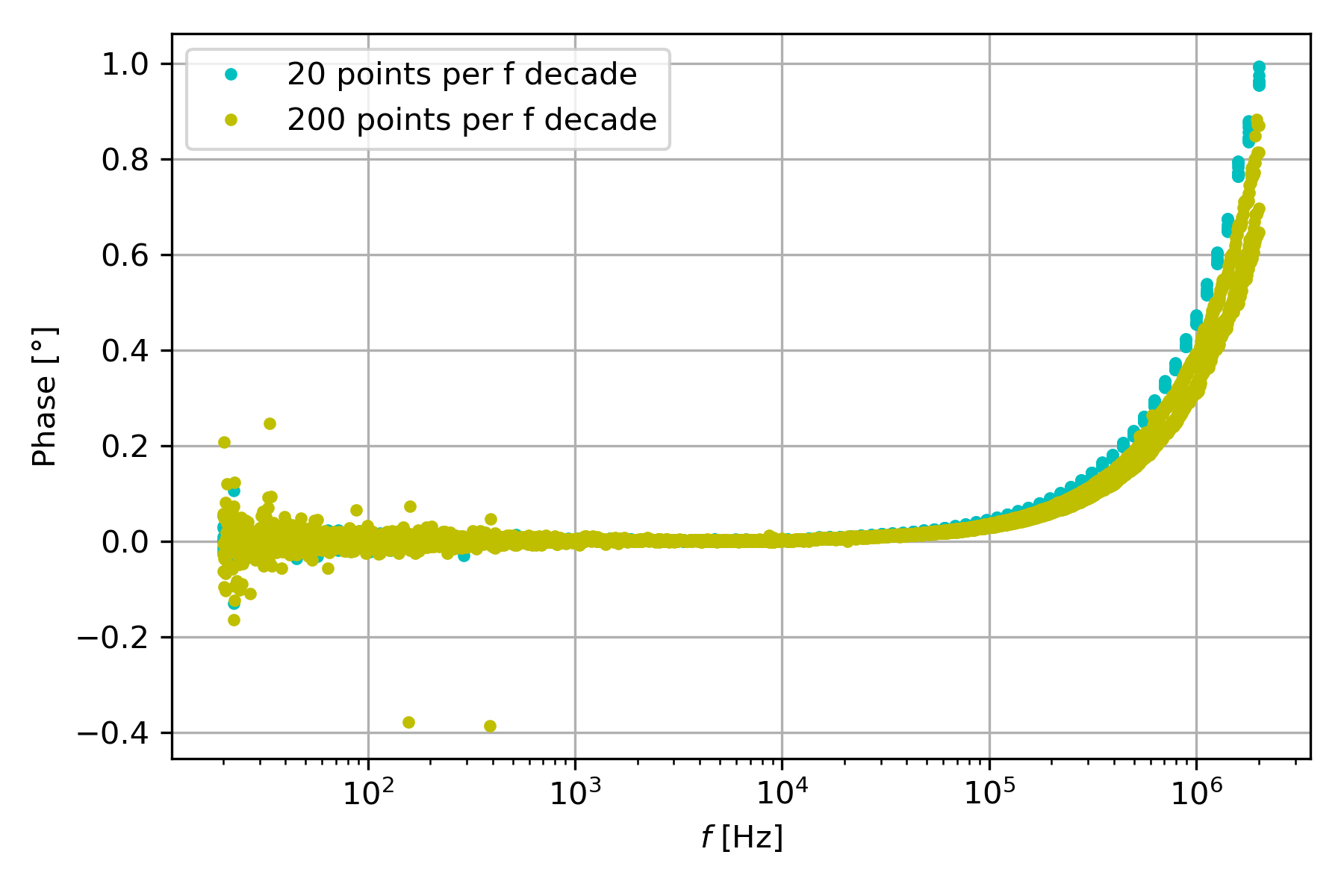}
        \caption{\label{fig:FS8} Two sequences of four \textit{f} sweeps (between 20 Hz and 2 MHz, setting the bias voltage at 0 V and the AC amplitude at 20 mV). The first sequence increases \textit{f} at a rate of 20 points per decade, while the second records 200 points per decade.}     
\end{figure}

\noindent By running two sweeps, one with ten times more points per frequency decade (meaning that the measurement is ten times longer in duration) and comparing them (see Fig. S\ref{fig:FS8}), we conclude this is independent of the elapsed time of the total measurement. This implies that the effect is genuinely dependent on the frequency value of the AC signal. Whether this is the reflection of a self-induction within the AgNWN or a spurious artifact is still unclear, mainly due to the magnitude of the effect, which prevents us from confidently fitting its associated Nyquist plot.  

%\newpage
%\printbibliography

\end{document}